\documentclass[fleqn,usenatbib]{mnras}
\usepackage{amsbsy}
\usepackage{amssymb}
\usepackage{amsmath}
\usepackage{graphicx}
\usepackage{xcolor}
\usepackage[caption = false]{subfig}
\usepackage [english]{babel}
\usepackage [autostyle, english = american]{csquotes}
\usepackage{natbib}
\MakeOuterQuote{"}

\title [Detecting superfluid transition] 
{Detecting superfluid transition in the pulsar core}
\author [Bagchi, Layek, Saini, Sarkar, Srivastava and Godaba Venkata]
{Partha Bagchi $^1$ \thanks{E-mail: parphy@niser.ac.in, parphy85@gmail.com}, 
Biswanath Layek $^2$ \thanks{E-mail: layek@pilani.bits-pilani.ac.in}, 
Dheeraj Saini $^3$ \thanks{E-mail: 23pph001@lnmiit.ac.in},
Anjishnu Sarkar $^3$ \thanks{E-mail: anjishnu@lnmiit.ac.in}, 
\newauthor
Ajit M. Srivastava $^4$ \thanks{E-mail: ajit@iopb.res.in} and 
Deepthi Godaba Venkata $^2$ \thanks{E-mail: 
p20210075@pilani.bits-pilani.ac.in}\\
$^1$ School of Physical Sciences, National Institute of Science Education 
and Research, Bhubaneswar, India.\\
$^2$ Department of Physics, Birla Institute of Technology and Science, Pilani-333031, India. \\
$^3$ Physics Department, The LNM Institute of Information Technology, Jaipur-302031, India. \\
$^4$ Institute of Physics, Sachivalaya Marg, Bhubaneswar-751005, India.} 
\date{}
\pubyear{}

\begin{document}
\label{firstpage}
\pagerange{\pageref{firstpage}--\pageref{lastpage}}
\maketitle

\begin{abstract}
It is believed that the core of a neutron star can be host to various
novel phases of matter, from nucleon superfluid phase to exotic high
baryon density quantum chromodynamics (QCD) phases. Different
observational signals for such phase transitions have been discussed in
the literature. Here, we point out a unique phenomenon associated with
phase transition to a superfluid phase, which may be the nucleon
superfluid phase or a phase like the CFL phase, allowing for superfluid
vortices. In any superfluid phase transition, a random network of
vortices forms via the so-called Kibble-Zurek mechanism, which
eventually mostly decays away, finally leaving primarily vortices
arising from the initial angular momentum of the core. This transient,
random vortex network can have a non-zero net angular momentum for the
superfluid component, which will generally be oriented in an arbitrary
direction. This is in contrast to the final vortices, which arise from
initial rotation and hence have the initial angular momentum of the
neutron star. The angular momentum of the random vortex network is
balanced by an equal and opposite angular momentum in the normal fluid
due to the conservation of angular momentum, thereby imparting an
arbitrarily oriented angular momentum component to the outer shell of
the neutron star. This will affect the pulse timing and pulse profile of
a pulsar. These changes in the pulses will decay away in a
characteristic manner such that the random vortex network decays,
obeying specific scaling laws leading to universal features for the
detection of superfluid transitions occurring in a pulsar core.
\end{abstract}


\section{Introduction}
\label{section:sec1}

Physics of neutron stars has been a very exciting area of research,
with their cores having matter at extreme baryon density. There has been
an explosion of interest in neutron stars with the direct detection
of gravitational waves by LIGO/Virgo coming from binary neutron star
(BNS) merger events. The properties of neutron star cores directly
affect gravitational waveform, especially during the last stages of
merger. There have been numerous investigations probing
the possibility of detecting such effects in the presence of high baryon 
density QCD phases, e.g., the deconfined QCD phase in the neutron star cores 
\citep{PhysRevLett.122.061102,PhysRevLett.122.061101} (see
also Ref. \citep{Demircik:2022uol,Radice:2016rys}).
However, for a nucleonic superfluid phase, there is convincing evidence of such a phase in 
the neutron star's inner crust and core. The existence of such a superfluid 
phase in the neutron star interior was first hypothesized by \cite{1959NucPh..13..655M}.
It was subsequently proposed that such a superfluid phase may be
responsible for the observed decrease in the slow-down rate
of the Vela pulsar in its post-glitch relaxation stage 
\citep{1969Natur.224..673B}. Superfluid phase is now believed to provide a natural 
explanation of pulsar glitches which can  arise from the de-pinning of superfluid vortices 
\citep{Anderson:1975zze,1976ApJ...203..213R}.
Nucleon superfluidity inside the neutron star's core also directly affects the neutron 
star cooling. The observed rapid cooling of the neutron star in 
Cassiopeia A (Cas A) supports the hypothesis of nucleon superfluid phase in the neutron star's 
core \citep{Shternin_2022}.  Note that after the discovery of the 
neutron star in Cas A \citep{Hughes_2000}, this star is found to be among 
a few isolated neutron stars with a well-determined age and a reliable surface 
temperature, thus allowing for modelling of its thermal evolution and the determination 
of its interior properties. \cite{PhysRevLett.106.081101} and \cite{Shternin_2011}
claimed that enhanced neutrino emission caused by the breaking and formation of neutron Cooper pairs 
in the $^3P_2$ channel is responsible for the rapid cooling of the neutron star in Cas A.  

Note that, superfluidity in condensed matter systems has been observed for more than hundred years
(e.g. in $^4$He, and later in $^3$He). Quantized vortices in superfluid
helium have also been routinely studied in laboratory experiments for a long 
time. Regarding superfluid phase arising from Cooper pairing of nucleons, the 
experimental evidence from laboratory experiments comes from the effects of 
pairing correlations in finite nuclei where it is manifested in the odd-A - even-A 
staggering of the binding energies \citep{Sedrakian_2019}. 
Such Cooper pairing of nucleons in the interior of neutron stars 
can also affect the cooling rate of neutron stars \citep{Shternin_2022,PhysRevLett.106.081101}. 
One of the observational (indirect) evidence for superfluid vortices in a 
nucleonic superfluid (arising from Cooper pairing of nucleons) has been in terms of pulsar glitches
(see the recent work \citep{PhysRevLett.132.181001} for further evidence of superfluid vortices
in the neutron star interior).

In this paper we will provide a novel phenomenon occurring in a neutron
star core which undergoes a superfluid phase transition. Our discussion
will only utilize the superfluid nature of the transition, and hence
will equally apply to conventional nucleon superfluid phase, as well as
the exotic extreme baryon density QCD phases such as the color-flavor
locked (CFL) phase which is also expected to be a superfluid phase
of the diquark condensate.  There have been many investigations of
the effects of phase transitions in the neutron star core, and its
observational signatures. Some of the signals are, effects on the
cooling rate of neutron stars (see, e.g \citep{PhysRevLett.106.081101} 
for nucleon superfluid case and \citep{Alford:2007xm,Rajagopal:2001ngu} for exotic
high baryon density QCD phases),
change in the moment of inertia of the core due to change in the free
energy of new phase \citep{Heiselberg:1998vh}, and the effects of
density fluctuations produced during a phase transition on the pulsar
rotational dynamics \citep{Bagchi:2015tna}.  In the investigations
of effects of phase transition induced density fluctuations, some of
us had also discussed specific case of superfluid phase transitions,
by focusing on the random density fluctuations produced by superfluid
vortex network \citep{Bagchi:2015tna}. However, the fact that superfluid
vortices involve fluid circulation was not utilized in those discussions,
only the effects on the energy density fluctuations was considered. Here
we focus on the fluid circulation associated with the vortex network
and show that it leads to specific observational signals.

Our discussion will be specific to a pulsar undergoing a superfluid phase
transition, which can be the nucleon superfluid phase, or a phase like CFL
phase, allowing for superfluid vortices. It is well established that in a
superfluid phase transition, a random network of vortices forms, via the
so called Kibble-Zurek mechanism \citep{Kibble:1976sj,Zurek:1996sj}. This
has been  seen in laboratory experiments and is well understood using
a picture of correlation domain formation during a phase transition. As
the phase of the superfluid condensate randomly varies from one domain
to another, it leads to formation of vortices at junctions of various
domains around which the phase winds non-trivially. The resulting
vortex network shows universal properties, and decays away with specific
scaling exponent \citep{Brandenberger:1993by, PhysRevLett.126.185302}.
This is the picture in the absence of any initial rotation present for
the matter undergoing superfluid transition. In the presence of initial
rotation, the Kibble-Zurek mechanism has to be modified, as apart from
the random vortex network, which are randomly oriented, extra vortices
need to be created to carry the initial net angular momentum of the
fluid. This modified mechanism was proposed by some of us in Ref.
\citep{Dave:2018fym}, and we will incorporate these considerations in
discussing formation of vortex network here.

The random vortex network  formed by the Kibble-Zurek mechanism (with
due consideration of initial rotation) will eventually mostly decay
away, finally leaving primarily vortices arising from initial angular
momentum of the neutron star. As vortices in this  random vortex network
are randomly oriented, in general there can be a non-zero net angular
momentum for the superfluid component, which can be oriented in an
arbitrary direction. Important thing is that the direction of this net
angular momentum will be completely independent of the original angular
momentum of the neutron star. Along with this random vortex network there
will also be extra vortices  which arise from initial rotation. These
extra vortices will carry the initial angular momentum of the neutron
star.  \citep{Das:2016cfg,Dave:2018fym} had pointed out an important
aspect of the formation of the random vortex network, which needs to be
incorporated in the discussions of the Kibble-Zurek mechanism, especially
in relation to experimental observations of fluid circulation. This
involves implementation of  angular momentum conservation, arising from
strict local linear momentum conservation, for the random vortex network
formation. We will discuss it below in detail. Basically one finds that
the angular momentum carried by the random superfluid vortex network
has to be balanced by an equal and opposite angular momentum in the
normal fluid. The end result is that the arbitrarily oriented angular
momentum component of the vortex network is finally transferred (with
opposite direction) to the outer shell of neutron star. This extra
angular momentum, though very tiny, will  induce a wobbling of the
neutron star, in addition to affecting its spin. This can be detected
for pulsars with high precision measurements of the pulse timings and
the pulse profile changes.

An important aspect of this signal, which can
distinguish it from other physical effects, arises from the fact that the
random vortex network decays obeying specific scaling laws. (This will
occur as vortices and anti-vortices annihilate, and also as vortices
pointing in different directions re-orient.)  Thus the perturbations
in the pulse timings and the pulse profile changes will also decay in
very specific manner consistent with those scaling laws. Such universal
features in observed pulses may give robust signals of superfluid phase
transitions occurring in pulsar cores. (In this context we 
mention that, although there is a strong evidence that the neutron star 
in Cassiopeia A may be undergoing superfluid phase transition 
\citep{Shternin_2022,PhysRevLett.106.081101}, it will be difficult to
test our proposed mechanism for this case as there has not been any 
detection of pulses from this  neutron star \citep{Ho:2009mm}.
A pulsar at similar stage of evolution will be an ideal candidate for 
testing the predictions of the mechanism proposed here.)

The paper is organized in the following manner. In section
\ref{section:sec2}, we discuss basic physics of superfluid phases
expected in the neutron star core.  Here we will discuss conventional
nucleon superfluidity as well as the superfluidity expected in extreme
baryon density QCD phases with (colored) diquark condensate.  In Section
\ref{section:sec3} we will discuss the Kibble-Zurek mechanism for the
formation of random vortex in a superfluid transition. Here we will also
discuss the considerations from Ref. \citep{Das:2016cfg,Dave:2018fym}
for the modifications in Kibble-Zurek mechanism needed to account for
extra vortices needed for systems with initial rotation, e.g. rotating
neutron star.  In Section \ref{section:sec4}, we will discuss estimates
of net, randomly oriented, angular momentum arising from the random
vortex formed during superfluid transition. Section \ref{section:sec5}
presents discussion of observational consequences of this extra, randomly
oriented, angular momentum on pulsar timing as well as pulse profile.
We conclude in section \ref{section:sec6} with discussion of various
uncertainties in our estimates, as well as the strengths of these
predictions in terms of universal features of these estimates.

\section{Superfluid phases in the core of a neutron star}
\label{section:sec2}
 
Here we will discuss basic physics of superfluid phases expected in the
neutron star core. We first discuss conventional nucleon superfluidity,
basic picture and estimates of relevant correlation length etc. Next we
will discuss basic physics of color superconducting QCD phases expected
to occur at extreme baryon density. Some of these phases, such as the
color flavor locked (CFL) phase, display superfluidity and we will
present model estimates of correlation length for this case.

\subsection{Nucleon Superfluidity}

The theoretical studies of neutron star physics for the last few decades,
supplemented with pulsar glitch data \citep{Manchester:2017ykr}, led to
the belief in the existence of neutron superfluidity in the interior of
neutron stars \citep{1969Natur.224..673B}. Below the outer crust region 
lies a few hundred 
meters thick inner crust of mass density $0.002 \rho_0
\le \rho \le 0.8 \rho_0$ ($\rho_0 = 2.8 \times 10^{14}~\text{g cm}^{-3}$
is the saturation density of nuclear matter.), which is believed to host
the neutron superfluidity in the $^1S_0$ BCS pairing state. The core of
about 10 km radius with density $\rho \ge \rho_0$ may contain neutron
superfluidity in $^3P_2$ state and a proton superconductor in $^1S_0$
state. Depending on the density of the core region, as we discussed
earlier, even more exotic QCD phases can appear in the superfluid form,
viz., CFL, LOFF and 2SC color superconductivity, etc. 
The presence of neutron superfluidity in the interior of the
neutron star is one of the main ingredients of the proposed superfluid-vortex 
model for pulsar glitches \citep{Anderson:1975zze,1976ApJ...203..213R}. The basic 
idea of this model is the sharing of excess angular momentum carried by the pinned superfluid 
vortices in the inner crust to the co-rotating outer (rigid) 
crust - core region of the star (see the reviews \citep{Haskell:2015jra, Antonopoulou:2022rpq} 
on pulsar glitches and various related issues). The success of the vortex model in 
explaining various features of pulsar glitches thus provides indirect evidence of 
superfluidity in the neutron star interior \citep{1969Natur.224..673B,Anderson:1975zze,1976ApJ...203..213R}.

Here, we briefly recall the theoretical arguments in favour of the
formation of nucleon superfluidity inside a neutron star, supplemented
with the values of relevant superfluid parameters obtained from various
model calculations. The effective nucleon-nucleon interaction is a
combination of short-range repulsion and a long-range attraction. As
a result, two distinct types of neutron superfluidity are expected
to occur in the interior. A few femtometer inter-particle distance in
the inner-crust region favours the long-range attractive interaction,
resulting in $^1S_0$ neutron superfluidity. At higher densities
($\ge \rho_0$), it's a competition between a short-range repulsion
and a long-range attraction, leading to $^3P_2$ pairing. 
The superfluid parameters, viz., the superfluid gap $\Delta$, the 
critical temperature $T_c$, etc., have been estimated  using
various many-body approaches (In this context, see the very informative 
reviews \citep{Dean_2003, Sedrakian_2019} on the quantitative understanding of 
superfluidity through the microscopic pairing theories in various nuclear systems, 
including the system of neutron-star interiors.). These parameters vary
with the baryon density. The various studies suggest that the superfluid
critical temperature in the core region lies in the (0.1 - 0.5) MeV range.
The cooling mechanism of neutron stars has been studied and
discussed extensively in the literature; see Refs. 
\citep{Yakovlev:2004iq,Yakovlev_2005,Page_2006}. The above studies suggest the 
star's typical interior temperature to be about $T \simeq 0.01$ 
MeV \citep{Yakovlev:2004iq}, which is well below the critical temperature and 
relatively uniform in the inner region of the star because of the high thermal 
conductivity of the degenerate quantum liquid.
Though our discussion will 
relate to vortex formation during the superfluid phase transition, at
temperatures near the critical temperature, our results will be reasonably
insensitive to the temperature dependent variation of
$\Delta$ in the superfluid phase. As we will see below, there is a very
weak dependence of number of extra vortices on the correlation
length (Eq.(\ref{eq:Nnet})), and hence on the temperature (Eq.(\ref{eq:xi})). 
Also, the observable effects of angular momentum from extra vortices will 
be significant only when the magnitude of the order parameter is a reasonable
fraction of its zero temperature value, with the relevant time scale relating to 
the coarsening of string network. Thus, in our 
order of magnitude estimates, we will ignore the temperature dependence of 
$\Delta$, and use its value at $T = 0$ as a typical value.

Here, we explore the effect on pulse modulation due to the formation
of random vortex networks in the core, resulting from the normal to
superfluid phase transition. This effect is determined by the number of
net vortices, which depends on the coherence length $\xi$ of the relevant
thermodynamic phases. Estimates of the coherence length in BCS theory
give a range of values for $\xi$ depending on the Fermi momentum $k_f$
of neutrons. For example, the authors of Ref. \citep{Gezerlis:2014efa}
(see also, Ref. \citep{Elgaroy:2001rg}) estimated the value of $\xi$
in the range of a few tens fm to order 100 fm, as the neutron's Fermi
momentum varies from 0.1 fm$^{-1}$ to 0.8 fm$^{-1}$. We will use a sample
value of $\xi = 100$ fm for our purpose.  We will see below that our
results have a very weak dependence on the value of $\xi$.

\subsection{Superfluidity in Color Superconducting Phases of QCD}

There are suggestions that the core of a neutron star, at very
high baryon chemical potential with low temperature, may host
various exotic thermodynamic phases of QCD \citep{Rajagopal:2000wf,
Alford:2001,Alford:2007xm}. These suggestions follow from the realization
that at very high values of baryon chemical potential, the physics should
be governed by the low energy excitations near the Fermi level. In this
case, the dynamics at the Fermi surface at $T \simeq 0$ is governed
by the quarks and the gluon-mediated quark-quark interactions. The CFL
phase may occur at ultra-high quark chemical potential of order 500 MeV,
where $u,d,$ and $s$ quarks can be treated massless.  As suggested in the
literature, this phase can occur if the NS core achieves an extreme mass
density. For CFL phase, the attractive interaction between the quarks
in (color antisymmetric) $3^*$ channel causes Fermi surface instability
favouring diquarks BCS pairing, leading to the {\it color superconducting
phase}. With color antisymmetric ($3^*$ channel), spin antisymmetric
(for $^1S_0$ pairing), the condensate should be flavor antisymmetric
with the structure,

\begin{equation}
\langle q_i^\alpha q_j^\beta \rangle \sim \Delta_{CFL} (\delta^\alpha_i 
	\delta^\beta_j - \delta^\alpha_j \delta^\beta_i) = \Delta_{CFL}
	\epsilon^{\alpha\beta n} \epsilon_{ijn}
\label{eq:qiqj}    
\end{equation}

where $\alpha\beta$ are flavor indices and $i j$ are color indices.
Note that the condensate is invariant under equal and opposite color and
(vector) flavor rotations. Hence, it is called as color-flavor locked
phase.  It leads to the following spontaneous symmetry breaking pattern,

\begin{equation}
	SU(3)_{color} \times SU(3)_L \times SU(3)_R \times U(1)_B
	\rightarrow SU(3)_{C+L+R} \times Z_2
\label{eq:symbreak}    
\end{equation}

Thus, the $SU(3)_C$ color symmetry of QCD, along with three flavors
chiral symmetry, is spontaneously broken. Fundamental group of the vacuum
manifold here is $Z$, giving rise to vortices. Superfluid nature arises
from the spontaneous breaking of $U(1)_B$.  Depending on the relative
mass difference between $u$, $d$ and $s$ quarks, other interesting phases
at high baryon chemical potential may also appear in the core, namely,
the $2SC$ phase for two light $u$ and $d$ quarks, or the LOFF phase, when
chemical potential is not too large compared to the strange quark mass.

Observational signatures of these phases have been discussed in the
literature \citep{Alford:2007xm,Rajagopal:2001ngu}. For example,
the central core with  CFL phase leads to suppressed cooling by
neutrino emission and has smaller specific heat. Thus, the outer
layer in the standard nucleonic phase will dominate NS's total
heat capacity and neutrino emission with the CFL core. From the
perspective of possible signatures of the exotic QCD phases, some of
us in Ref. \citep{Bagchi:2015tna, Srivastava:2017itj,Bagchi:2021etv}
suggested that the above symmetry breaking phase transitions may cause
density fluctuations in the core by forming topological defects, leading
to a transient change of moment of inertia (MI) tensors components. 
\citep{Bagchi:2015tna} have shown that the change of diagonal components of 
the MI tensor may lead to the change of the spin frequency of the
pulsars and may be responsible for glitches and/or anti-glitches.
The fluctuations being random, the generation of quadrupole moments
may lead to the emission of gravitational waves. In the subsequent
work \citep{Bagchi:2021etv}, the authors have studied the effect of
the development of the non-zero off-diagonal components of the MI tensor
on pulse profile. Following the spirit of those works 
\citep{Bagchi:2015tna,Bagchi:2021etv}, here we shall explore the possibility 
of probing the superfluid phases (neutron superfluid and CFL) through the 
observational effects on pulsars due to the formation of random topological 
vortices in the core implementing the suggested new mechanism of random 
vortex formation in Ref. \citep{Das:2016cfg,Dave:2018fym}.

For the estimate of coherence length $\xi$, we shall follow the Ref.
\citep{PhysRevD.66.014015}, where the coherence length has been estimated
for the CFL phase within the Ginzburg-Landau theory in weak coupling
limit as,
  
\begin{equation}
\xi = 0.26~ \left(\frac{100~\text {MeV} }{T_c}\right) 
\left(1-\frac{T}{T_c}\right)^{-1/2}~ \textrm{fm} .
\label{eq:xi}
\end{equation}

The critical temperature $T_c$ for CFL transition is related
to the CFL gap $\Delta_0$ at $T = 0$ through $T_c \simeq 0.57
\Delta_0$. The value of $\Delta_0$ lies in the range (40 - 90) MeV (See
Refs. \citep{Berges:1998rc, Evans:1998ek} for various issues related to
estimating the superfluid parameters.), resulting in $T_c \simeq (20
- 50)$ MeV. The critical temperature $T_C$ for CFL transition is too
high compared to the typical star's temperature, favouring such phase
at extremely high baryon density. The above values of $T_c$ and $T$
provide the coherence length $\xi$ in the (0.5 - 1) fm range.

\section{Formation of random vortex network in a superfluid transition}
\label{section:sec3}

Here we will discuss how a superfluid phase transition leads to
formation of a dense random vortex network. This random vortex
network forms in addition to vortices needed to account for any
initial rotation of the system.  The formation of random network
of vortices arises from the so called Kibble-Zurek mechanism
\citep{Kibble:1976sj,Kibble:1980mv,Zurek:1996sj}.  However, this
mechanism needs modification to introduce a bias for the formation
of extra vortices for the case when the system has initial rotation
\citep{Das:2016cfg,Dave:2018fym}. We will discuss basic physics of these
modifications needed. However, for the case at hand, we will find that the
density of extra vortices arising from initial rotation is many orders
of magnitude smaller than the random vortex network density arising
from the usual Kibble-Zurek mechanism. Thus, for this case, one can get
reasonable estimates by ignoring these modifications. However, we will
discuss an important aspect of the formation of random vortex network
from Refs. \citep{Das:2016cfg,Dave:2018fym}
which needs to be incorporated in the discussions of the Kibble-Zurek
mechanism for prediction of observed effects of fluid circulation.
This will involve implementation of local linear momentum conservation,
for the random vortex network formation, and will lead to the conclusion
that the flow generated for the superfluid component during formation
of vortices be balanced by an equal and opposite momentum in the normal
fluid.  The end result will be that the arbitrarily oriented angular
momentum component of the vortex network is finally transferred (with
opposite direction) to the outer shell of neutron star.

Superfluid vortex is one example of topological defects. Topological
defects arise in a wide range of systems. There are numerous examples
of topological defects in condensed matter systems. It is also expected
that certain topological defects arise in symmetry breaking transitions
in the early universe.  The first detailed theory of formation of
topological defects via a domain structure arising during a phase
transition was proposed by \citep{Kibble:1976sj,Kibble:1980mv} in the
context of early universe. It was subsequently realized that this {\it
Kibble mechanism} applies equally well to any symmetry breaking transition
\citep{Zurek:1996sj} (see also \citep{Gupta:2010pp,Gupta:2011ag,Mohapatra:2012ck}).  
This has
allowed the possibility of testing the predictions of Kibble mechanism in
various condensed matter systems, see Refs. \citep{1994Natur.368..315H,
Ruutu:1995qz, PhysRevLett.81.3703, PhysRevLett.84.4966,Volovik:1996qw,
PhysRevLett.84.4966,PhysRevLett.91.197001,arxiv.condmat.0312082, 
PhysRevLett.85.3452,Rudaz:1994dqy, PhysRevA.45.R2169, PhysRevE.47.3343, 
Chuang:1991zz,Bowick:1992rz,PhysRevD.69.103525,PhysRevLett.83.5030}. 
There are important issues for defect formation in continuous transitions due 
to critical slowing down \citep{Zurek:1996sj}. The Kibble-Zurek mechanism
incorporates these aspects and leads to specific predictions of the
dependence of  defect densities on the time scale of the phase transition.

An important aspect of the theory proposed by Kibble is that
the basic mechanism has many universal predictions making it
possible to use condensed matter experiments to carry out rigorous
experimental tests of the predictions made for cosmic defects
\citep{Bowick:1992rz,PhysRevD.69.103525, PhysRevLett.83.5030} in
laboratory experiments. One of such universal predictions relates
to the correlation between the formation of defects and anti-defects
\citep{PhysRevLett.83.5030}, and we will be using this specifically in
our calculations below.

The crucial element of the theory proposed by Kibble is the recognition
that phase transitions lead to formation of a sort of domain structure
with the order parameter field varying randomly from one domain to
another. Individual domains correspond to the correlation regions
where order parameter field is taken to be roughly uniform. Second
important input in this  theory is the assumption that the order
parameter field between two adjacent domains varies along the shortest
path on the vacuum manifold.  This is called as {\it the geodesic
rule}. Its validity for different cases, such as those involving gauge
symmetries, and possible violations in certain specific situations
which are dominated by fluctuations has been discussed in the literature
\citep{Rudaz:1992wy,PhysRevD.55.3824}.  With these two physical inputs,
one gets a geometrical picture for the physical region undergoing phase
transition. One can then use straightforward topological arguments to
calculate the probability of formation of defects and anti-defects at
different junctions of domains. The probability of defect formation
in this theory is calculated {\it per correlation domain} and it is
a universal prediction, depending only on the underlying symmetries
and space dimensions. This universality has allowed experimental test
of the prediction of defect density in liquid crystal experiments
\citep{Bowick:1992rz} (see also \citep{PhysRevD.69.103525,PhysRevLett.83.5030}) 
for a first order transition case where correlation domains could be directly
identified as bubbles of the nematic phase nucleating in the background
of isotropic phase.  For a continuous transition, effects of critical
slowing down introduce dependence of relevant correlation length on
the rate of phase transition \citep{Zurek:1996sj}. The Kibble-Zurek
mechanism incorporates these considerations in prediction of defect
density \citep{Zurek:1996sj}.

In the following, we will discuss basic elements of this theory for the
specific case of spontaneous breaking of global U(1) symmetry. This is
the case for the superfluid transition for nucleon superfluidity, as well
as for superfluidity in the CFL phase. Superfluid component is 
characterized by the condensate wave function, $\Psi=\Psi_0 e^{i\phi}$, 
where $\Psi_0^2$ gives number density of superfluid component and the
phase $\phi$ is related to the superfluid velocity $\vec{v_s}$,

\begin{equation}
\vec{v_s}=\frac{\hbar}{m}\vec{\nabla} \phi
\label{eq:vs}
\end{equation}

Here $m$ is the mass of the atomic unit which is condensing, for
superfluid $^4$He, it will be mass of $^4$He atom, for our case of neutron
superfluidity, it will be  case the mass of Cooper pair of neutrons,
hence $m = 2m_N$, $m_N$ being mass of neutron. Superfluid vortices arise
where the phase $\phi$ winds non-trivially, we may call vortex as those
points around which net variation of $\phi$ is $+2\pi$ with anti-vortex
corresponding to $-2\pi$ variation of $\phi$. (Similarly, for higher
windings with multiples of $2\pi$.)  The order parameter space in this
case is a circle $S^1$. According to Kibble's theory, defects form due to
the domain structure arising in the phase transition. The domains here
will be characterized by roughly uniform $\phi$ which varies randomly
from one domain to another. In accordance to the geodesic rule, $\phi$
will vary with least gradient in between adjacent domains. By considering
the probability of finding a non-zero ($\pm 1$) winding around a junction
of three domains, one can show \citep{Bowick:1992rz,PhysRevD.69.103525}
that the probability of vortex/antivortex formation per domain, in
two space dimensions, is equal to 1/4. This can be straightforwardly
generalized to three space dimensions where one gets vortex line
(topological string) defects. Simulations show that the average number
of strings per correlation domain is 0.88 \citep{Vachaspati:1984dz}.
Interestingly, the statistical properties of initial string network are
universal, with strings basically forming Brownian trajectories with
persistence length of the correlation length. There are large number of
string loops, while many strings stretch across the system.

Vortex here corresponds to positive circulation of fluid, with anti-vortex
corresponding to opposite circulation. As we are interested in net
angular momentum, we need to find net vortex $-$ antivortex number
which are ending at a given surface. This will give net angular momentum
perpendicular to that surface. An important point here is that one needs
to know here the angle at which a given vortex or antivortex ends at the
surface.  That will give the direction of angular momentum carried by the
vortex/antivortex at that surface. We will be using primarily topological
arguments, which will only give probability of non-zero winding at the
surface. Detailed simulation like in Ref.\citep{Vachaspati:1984dz} can
give the information of the angle. However, this will only contribute
to a factor of order one.  As we will see, in our estimates, factors of
order one will be completely irrelevant, hence we will not worry about
this issue.

One may expect that if $N$ is the net number of vortices plus antivortices
ending at a surface, then the number of vortices minus number of
antivortices, $\Delta N$ will typically be of order $\sqrt{N}$. This,
however, is not true. The underlying domain structure induces certain
correlation between vortices and antivortices \citep{PhysRevLett.83.5030},
due to which $\Delta N$ is strongly suppressed. In fact, domain
structure directly allows us to calculate $\Delta N$ for any given
area. Consider a 2-dimensional surface $S$ which has area $A$, and
perimeter length $L$. The perimeter will consist of $L/\xi$ number of
correlation domains, where $\xi$ is the correlation length.  With $\phi$
varying randomly from one domain to another, we essentially have here a
random walk problem for $\phi$ where the average step size for $\phi$
is $\pi/2$ (as the largest variation in $\phi$ step is $\pi$ and the
shortest is zero). We can then conclude that the net winding of $\phi$,
as we go around the perimeter $L$, (which will give the net vortex minus
anti-vortex number $\Delta N$ enclosed within $L$), will be distributed
about zero with a typical width $\sigma = \frac{\pi/2}{2\pi}
\sqrt{\frac{L}{\xi}} = \frac{1}{4}\sqrt{\frac{L}{\xi}}$.  The total number of
vortices plus anti-vortices $N$ is proportional to the area $A$ (with a
fixed probability of vortex/antivortex per domain, which is equal to 1/4
for 2 space dimensions for this case). Thus, we conclude that $\sigma
\propto N^{1/4}$ which is in complete contrast to the naive expectation
of the Poisson distribution for $\Delta N$ with width $\sqrt{N}$. We
thus get a scaling relation for $\sigma$:

\begin{equation}
\sigma = C N^\nu
\label{eq:sigma}
\end{equation}

The exponent $\nu$ is universal here with a value of $\nu = 1/4$ for the
case of vortices with spontaneous breaking of $U(1)$ symmetry. We again
emphasize, this highly non-trivial prediction arises from the underlying
domain structure in this picture of vortex formation, and is in complete
contrast to naive expectation of $\nu = 1/2$ for random sprinkling of
vortices and anti-vortices. The constant $C$ is not universal and depends
on details like shapes of elementary correlation domain etc. Typically
its value if of order 1 (e.g. $C = 0.57$ and 0.71 for triangular and square
domains respectively), and as mentioned above, factors of order 1
will be completely irrelevant in our present case, as we will see below.

The prediction of the distribution of $\Delta N$ being centered at zero
(and with $\sigma$) is for the case when vortices form in a system
which has no initial rotation. For a rotating system, a network
of vortices (vortex lattice) has to finally form which carries the
initial angular momentum.  These extra vortices will be in addition
to the vortex network formed via the Kibble-Zurek mechanism. Thus the
basic picture of Kibble-Zurek mechanism has to be modified so that
extra vortices are produced during the transition in accordance to
the initial angular momentum of the system. A detailed investigation
of this was carried out in Ref.\citep{Das:2016cfg,Dave:2018fym} where
it was shown that it requires modification of both crucial elements
of the Kibble's basic picture of defect formation via domains. Order
parameter within a correlation domain can no longer be assumed to be
roughly constant, indeed $\phi$ must vary everywhere in accordance to the
initial rotation velocity of the fluid (which becomes rotation velocity
of the superfluid). Further, the geodesic rule also gets modified as
$\phi$ variation within correlation domains has a non-trivial space
dependence. This affects the entire distribution of defects, and hence
also the distribution of $\Delta N$ (though the effect on $\sigma$
is small). We refer the reader to Ref.\citep{Das:2016cfg,Dave:2018fym}
for these details.  However, in the present case, typical separation
between vortices/anti-vortices formed during transition will be of order
the correlation length, ranging from 1 fm to 100 fm for CFL vortices or
neutron superfluid vortices respectively, whereas typical inter-vortex
separation arising from initial rotation of neutron star is of order 0.1
mm. This implies that number of extra vortices will be many orders of
magnitude smaller than the number of vortices/antivortices formed during
the phase transition (about 18 to 20 orders of magnitude smaller). Thus,
we can safely neglect effect of initial rotation on the basic mechanism of
vortex formation here, and will continue to use the estimate of $\sigma$
as given in Eq.(\ref{eq:sigma}). We mention here 
that the modification of the Kibble mechanism due to effects of initial 
rotation becomes important only when order parameter variation in each
domain is significantly modified. This happens only when the net vortex 
number (number of vortices minus number of antivortices) becomes a reasonable 
fraction of the total number of vortices (vortices plus antivortices), which is
not the case here (see Ref.\citep{Dave:2018fym} for details).

However, there is one important aspect of superfluid vortex formation
which needs to be incorporated in discussions of Kibble-Zurek mechanism,
especially in relation to the experimental observations of fluid flow
development during the transition. (See, \citep{Das:2016cfg,Dave:2018fym}
for a detailed discussion of this).  During phase transition, the
spontaneous generation of flow of the superfluid due to vortex formation
in some region means that some fraction of neutrons (or diquarks for the
CFL case) form the superfluid condensate during the transition and develop
momentum due to the non-zero gradient of the phase of the condensate. The
remaining neutrons (or quarks, which form the normal component of fluid
in the two-fluid picture) will then develop opposite linear momentum
so that the linear momentum is local conserved. This implies that the
normal fluid around each vortex formed should develop a flow which is
exactly the same but opposite to the profile of the superfluid velocity
of the vortex, depending on relative fraction of the normal fluid and
the superfluid. (As emphasized in Refs.\citep{Das:2016cfg,Dave:2018fym},
these arguments show that during superfluid transition, as vortices
form, superflow and normal flow will have initial opposite directions,
so experimental detection will be complicated, unless one finds a way to
distinguish between normal flow and superflow, e.g. with flow evolution
due to viscous effects of the normal flow).

In our present case, the above discussion has very important implication. As 
the neutron star core undergoes superfluid transition, random vortex network
will form which will impart an arbitrarily oriented net angular momentum
to the superfluid of the core (in addition to the angular momentum coming from
the initial rotation of the core). However, the superfluid part is decoupled
from rest of matter in the neutron star, so this additional random angular
momentum does not transfer to the outer shell of the neutron star. (Note,
the situation is different for the initial angular momentum. Initially, the 
entire neutron star is rotating, including the outer shell, neglecting any
differential rotation. Even though the core becomes superfluid, and decouples
from rest of the matter, apart from the pinning of vortices, the shell
has same rotation velocity in the beginning.) However, in view of the
above arguments of local linear momentum conservation, we conclude that
the normal matter in the neutron star will develop an angular momentum which is
equal and opposite to the net randomly oriented angular momentum arising from
the vortex network formed during the transition. This will be transferred to
the shell and should be observable in perturbations of the pulses of the
pulsar. In the next section we will make estimate of this extra 
angular momentum.

\section{Estimates of extra angular momentum arising from the random 
vortex network}
\label{section:sec4}

The basic picture of random vortex network formation discussed above
can now be used to estimate any extra angular momentum arising from
this network.  We will take neutron star with radius of 10 km, and
will consider a spherical core region with radius $R_c$ which undergoes
superfluid phase transition.  $R_c$ can be as large as about 9 km for
nucleon superfluidity, while for extreme baryon density QCD phases
(e.g. CFL phase) it may be 5 km or less.  As the superfluid transition
proceeds, two sets of vortices will form.  One set will arise from the
initial rotation of the fluid, which will lead to a relatively dilute
system of vortices (with typical inter-vortex spacing of about 0.1
mm). The second vortex set is of interest to us, and it consists of
vortices/antivortices formed via the Kibble-Zurek mechanism.  This will
be an extremely dense network of vortex line defects which will have
initial shape of Brownian trajectories, with some ending at the surface
of the core region of radius $R_c$, while most will form closed loops
entirely enclosed within this region. (See, \citep{Vachaspati:1984dz} for
the universal features of the statistical distribution of such string
defects.)  Typical separation between vortex/antivortex line defects
will be of order of the correlation length $\xi$ for the superfluid
phase transition.  It is important to note here that the only relevant
detail of the specific model of superfluid phase transition here is the
correlation length $\xi$.  Everything else will be entirely determined
in terms of $\xi$. We will take sample values of $\xi \sim 1$ fm for the
superfluidity in the color superconducting QCD phase (e.g. CFL phase)
\citep{PhysRevD.66.014015}, and $\xi \sim 100$ fm for the conventional
neutron superfluid case. Again, we will see, that factors of order one
will be irrelevant to our estimates.

As discussed above, the density of vortices arising from the initial
rotation is many orders of magnitude smaller than the density
of vortices formed via the Kibble-Zurek mechanism. Therefore, we
will ignore any effects of rotation on the Kibble-Zurek mechanism
(which requires modifications for a rotating system as discussed in
Ref.\citep{Das:2016cfg,Dave:2018fym}). We further note that any vortex
loop which is entirely enclosed within the core region of size $R_c$
does not contribute to any net angular momentum for the core. If at all,
it can contribute to localized patterns in the fluid flow. However, any
vortex/antivortex ending at the surface of the superfluid core region
will contribute to a net angular momentum.  The total angular momentum of
the superfluid core region can be found by simply adding contributions
coming from all vortices and antivortices ending at different points on
the surface of this spherical core region with radius $R_c$.

As vortices and anti-vortices contribute to opposite fluid circulations,
clearly the quantity of interest to us is the net vortex number $\Delta N$
= vortex number $-$ anti-vortex number. As we explained in Section II,
$\Delta N$ is distributed about zero with typical width of $\sigma$.
$\sigma$ is given in Eq.(\ref{eq:sigma}) as calculated using the domain 
structure underlying the Kibble-Zurek mechanism. Again, we neglect any effect
of initial rotation on this calculation of $\Delta N$, and $\sigma$,
first due to extremely dilute system of vortices arising form initial
rotation, and secondly that modified Kibble-Zurek mechanism shows
very weak dependence of the exponent $\nu$ on the system rotation
\citep{Das:2016cfg,Dave:2018fym}.

To use Eq.(\ref{eq:sigma}) for estimating typical net vortex number $\Delta 
N$, we need to use a 2-dimensional region of area $A$. The relevant 
2-dimensional surface for us is the surface of the spherical core region with 
radius $R_c$. However, here we note a problem as this surface is a closed
manifold, forming a 2-sphere $S^2$. A closed manifold necessarily
introduces correlations between vortices and antivortices ending at the
surface, as every vortex ending at one point on this surface of $S^2$
has to have an exit point on this $S^2$. The estimate of $\sigma$ in
Eq.(\ref{eq:sigma}) does not incorporate such correlations. However, we will 
ignore this complication using the following justification. As vortex  strings
form typical Brownian trajectories, with persistence length of $\xi$
(which is in the range of 1 - 100 fm), it is extremely unlikely for a
given vortex to continue all the way to the opposite end of the surface
of $S^2$ which has a radius of about  5 - 9 km. Most often such a vortex
will bend around quickly and will end at some point on $S^2$, not too
far from the initial point, (where it will exit as an anti-vortex). Thus,
the surface of $S^2$ with a radius of 5-9 km can be viewed as consisting
of smaller patches (still, orders of magnitude larger than $\xi$), and
the estimate of $\sigma$ in Eq.(\ref{eq:sigma}) can be applied to each such 
patch, while ignoring the fact that all these patches are eventually glued to
form a closed manifold. (For a proper check of these arguments, 
one needs to carry out detailed simulation of string formation in a spherical 
region $S^2$, with open boundary conditions, which allow windings to form on  
the surface of this $S^2$, which extend to string defects in the interior 
region. This is non-trivial as typically one performs simulations with 
periodic, or closed (fixed), boundary conditions. We hope to carry out such 
a simulation in a future work.)

 It is important to note that this argument
cannot be used for the net vortex number which arises from the initial rotation
of the system. As we explained above, initially rotating system must ensure
that finally there is a required net vortex number (vortices minus antivortices)
in the direction of initial rotation (which will be ensured by a suitably
modified Kibble mechanism \citep{Dave:2018fym}). Thus, even with Brownian
trajectories of vortices with a very small step-size, correct number of 
net vortices will be ensured right in the beginning of vortex formation,
when the order parameter is settling down to its superfluid phase value.
The extra vortices we have calculated, arise at a later stage when superfluid
phase is established, and gradient energies arising with application of
geodesic rule have resulted in appropriate superfluid flow (with normal fluid
developing opposite flow, as explained earlier).

With these points clarified, rest of the calculations are straightforward.
Net number of correlation domains on the surface of $S^2$ with radius $R_c$
is given by:

\begin{equation}
N_{domains} \simeq \frac{4\pi R_c^2}{\xi^2}
\label{eq:Ndomains}
\end{equation}

Typical value of $\Delta N$ is characterized by $\sigma$ in 
Eq. (\ref{eq:sigma}). We thus estimate typical value of net vortex number as

\begin{equation}
\Delta N \sim \sigma = C N^{1/4} = C ~(4\pi)^{1/4} \sqrt{\frac{R_c}{\xi}} 
\label{eq:deltaN}
\end{equation}

We now note that these extra vortices will be roughly uniformly distributed
on the entire surface of $S^2$ which is the boundary of the superfluid core 
region. A vortex will contribute to positive angular momentum along
local perpendicular direction. (Again as we discussed earlier, any inclination 
of the angular momentum arising from the angle of the vortex string ending 
at the surface will contribute to factor of order 1 which will be immaterial
here.)  Thus, on the average, equal number of vortices will exit in positive
$x$ direction, and negative $x$ direction, similarly for $y$ and $z$ directions. 
Thus, one may expect roughly $\sigma/6$ number of vortices to typically exit
along a given direction, (with roughly equal number on average to exit
in the opposite direction, and same in other two directions).
However, this is only on average, there will be statistical fluctuations, and
in this case, distribution of these extra vortices will be expected to
follow Poisson distributions as the domain structure introduces no further
evident restrictions on how these extra vortices are distributed. (Validity of
this expectation needs to be checked by detailed numerical simulations).
We thus conclude that after all the cancellations of opposite circulations
(due to the extra vortices exiting in different directions), one will
expect a net number $N_{net}$ vortices to point along some specific 
direction, which should be typically of order

\begin{equation}
N_{net} \sim \sqrt{\frac{\sigma}{6}} = \sqrt{\frac{C}{6}} N^{1/8} =
	\sqrt{\frac{C}{6}} (4\pi)^{1/8} \left(\frac{R_c}{\xi}\right)^{1/4} 
\label{eq:Nnet}
\end{equation}

With $C$ being of order 1, $N_{net}$ is entirely determined by the
last factor of ${(R_c/\xi)}^{1/4}$. Further, core size of $R_c$ = 5
or 9 km also makes no difference. We take $R_c$ = 9 km. With that we
find $N_{net} \simeq 10^4$ for $\xi$ = 100 fm (for neutron superfluid case),
and $N_{net} \simeq 10^5$ for $\xi$ = 1 fm (for superfluidity in the
color superconducting case). Here we have ignored factors of order 1.

We now recall from the discussion at the end of previous section, that
this extra angular momentum of superfluid vortices has to be counter 
balanced by the generation of same, but opposite, angular momentum in 
the normal fluid component due to local linear momentum conservation. 
Thus, we conclude that during superfluid transition in a spherical
region of radius $R_c \simeq 5 - 9$ km, a net angular momentum
will be generated which will be pointing in an arbitrary direction, completely
uncorrelated to the original direction of rotation of the neutron star. 
The magnitude of this angular momentum will be equal to the angular
momentum carried by about $10^4$ to 10$^5$ superfluid vortices in this
spherical region, the two numbers corresponding to the neutron superfluid
case, and superfluidity in superconducting QCD phase, respectively.

\section{Observational effects, pulse timing and pulse profile changes}
\label{section:sec5}

We now estimate the value of the angular momentum arising from these net
vortices, pointing in an arbitrary direction at the surface of the
superfluid core of radius $R_c$. Quantization of the circulation around
a superfluid vortex means that the angular momentum of each Cooper pair
for a vortex (with unit circulation) is $\hbar$. Thus, net angular
momentum arising from $N_{net}$ number of vortices is

\begin{equation}
L_{vortex} = \hbar ~N_{net} ~ \left(\frac{M_{SF}}{m_{pair}}\right)
\label{eq:Lvortex}
\end{equation}

Here $M_{SF}$ is the mass of the superfluid component of the core. There
is a range of estimates of $M_{SF}$ in the literature 
\citep{Jones:2009zq,Sourie:2020ima} 
varying from few percent to almost 80 \% of the neutron star mass. For rough 
estimates, we will take $M_{SF} \sim M_{sun} \simeq 10^{33}$ grams.
$m_{pair}$ is the mass of the Cooper pair which will be equal to
2$m_{neutron}$ for neutron superfluidity, while $m_{pair} \simeq
m_{diquark} \simeq$ few hundred MeV for the color superconductivity
occurring at very high baryonic chemical potential. Again, for rough
estimates, we will take $m_{pair}$ = 2 GeV for both cases.  To represent
a general case, we consider the situation when $L_{vortex}$ points in a
direction transverse to the direction of the original angular momentum
$L_0$ which we take to be along the z axis. With this extra angular
momentum, the outer shell now has the net angular momentum pointing in
a direction which is tilted by an angle $\theta$ w.r.t the z axis. We
take $L_0$ to be

\begin{equation}
L_0 \simeq 10^{45} (\text{g~cm}^2) \frac{2\pi}{T_0} 
\label{eq:L0}
\end{equation}

where $T_0 (= \frac{2\pi}{\omega_0})$ is the initial time period (in
sec.) of the rotation of the neutron star. With this, the angle $\theta$
will be given by (for $L_{vortex} \ll L_0$)

\begin{equation}
\theta \simeq \frac{L_{vortex}}{L_0} = 
\frac{\hbar~N_{net} \left(\frac{M_{SF}}{m_{pair}}\right)}
{10^{45} (\text{g~cm}^2) \omega_0} 
\simeq N_{net}~ T_0 ~10^{-16}
\label{eq:theta}
\end{equation}

For millisecond pulsars, we get values of $\theta$ as $10^{-15}$ to $10^{-14}$
for the two cases of superfluidity. For slow rotating pulsar, with
$T_0 \sim 1$ sec. $\theta \simeq 10^{-12}$ to $10^{-11}$ for the two cases.

For observational effects, we consider specific case of the pulsar having
shape of an oblate spheroid, initially rotating about the principal
symmetry axis $z$. We take $I_{zz} > I_{xx} = I_{yy}$ with oblateness of
the pulsar characterized by $\eta = \frac{I_{zz} - I_{yy}}{I_{zz}}$. With
extra angular momentum leading to the tilting of net angular momentum
away from the principal symmetry axis $z$, pulsar will begin to wobble,
with the wobbling frequency $\Omega$ given by

\begin{equation}
\Omega \simeq \eta \omega_0
\label{eq:Omega}
\end{equation}

\noindent
where $\omega_0 = 2\pi/T_0$ is the initial angular frequency
of rotation. The structure of modulated pulses can be calculated
straightforwardly using Euler's equations (see, e.g. Ref.\citep{Bagchi:2021etv}).
The angular amplitude of wobbling will be given by

\begin{equation}
\theta_{wobble} \simeq \theta \frac{\omega_0}{\Omega} 
\simeq \frac{\theta}{\eta}
\label{eq:thetawobble}
\end{equation}

Neutron stars typically have very small values of $\eta$. Taking a sample 
value of $\eta \sim 10^{-6}$ we find angular amplitude of wobble to be about 
10$^{-5}$ to 10$^{-6}$ for the cases of CFL superfluidity and neutron 
superfluidity respectively. This should be detectable in observations of 
pulse modifications.

For the cases when $L_{vortex}$ is along the initial rotation axis  ($z$ 
axis), it will appear as a glitch, suddenly increasing the rotation frequency,
or decreasing it, which will then appear as anti-glitch. Note that
this gives the correct range of glitch amplitude as we will get

\begin{equation}
\frac{\delta \omega}{\omega} = \frac{L_{vortex}}{L_0} 
\label{eq:domegaByomega}
\end{equation}

This can range from $10^{-15}$ to 10$^{-11}$ as shown above. Further,
this extra angular momentum does not have to vanish at the end of the
coarsening of the vortex network.
Importantly, in this picture, one gets a unified explanation of glitches
and anti-glitches which correspond to $L_{vortex}$ being along the direction
of $L_0$, or opposite to it, respectively. Note that we
are not implying that this mechanism can provide explanation of all
glitches and anti-glitches. First, in our mechanism, glitches/anti-glitches
will occur in neutron stars which are undergoing superfluid transition.
However, glitches have been observed in very old pulsars with internal 
temperatures much lower than the superfluid critical temperature. Further,
glitches have been observed multiple times in certain pulsars, which will be
very hard to accommodate in the present scenario. Also, glitches with
much larger magnitudes have been observed, than the magnitudes estimated
here. Thus, as far as glitches/anti-glitches are concerned, the present 
mechanism may only be able to account for some of them, and one has
to invoke other very established mechanisms for most of these. The
present mechanism should be considered as providing a new phenomenon, and
its observational effects should have distinct features compared to
the other mechanism which are needed to explain a large range of glitches
and anti-glitches which have been observed.

An extremely important aspect of our model is that it predicts a universal
behavior of changes in the rotational dynamics of pulsar. $L_{vortex}$
arises from the random vortex network formed during the superfluid
phase transition.  The decay of this network follows universal scaling
laws. For example, for the 2-d case, vortex number scales as $t^{-\zeta}$
where $\zeta$ may be 1 or 1/3 depending on the nature of vortex-antivortex
interaction \citep{Brandenberger:1993by,PhysRevLett.126.185302}. The
observational effects of the random vortex network are all proportional
to $L_{vortex}$ which is determined by the net number of extra vortices
$N_{net}$ exiting the surface of $S^2$ in certain direction. 
Eq.(\ref{eq:Nnet}) shows that $N_{net}$ is proportional to $N^{1/8}$, with 
$N$ being the total number of vortices and anti-vortices. With $N$ decaying as
a power law, we expect that $L_{vortex}$ will decay as a power law
$t^{-\zeta/8}$. Note, the net vortex number here can change as vortices
exiting the surface of $S^2$ in different directions will re-orient
during evolution. Thus vortices on the surface of $S^2$ can slide and
can annihilate oppositely oriented vortices in different directions. In
general, due to curved surface of $S^2$, one should expect very few extra
vortices to survive, only those which stretch along the direction of
rotation of the star. As we mentioned, this will be consistent with the
behavior of glitches/anti-glitches where the initial angular momentum
is not fully restored.  During this coarsening of vortex network, all
observational effects, from wobbling frequency, to wobbling amplitude,
and the amplitudes of glitches/anti-glitches will be expected to decay
with this power law.  This will be a unique distinguishing signal of
the applicability of our model, and a direct signal of superfluid phase
transition occurring inside a pulsar core.

\section{Conclusion}
\label{section:sec6}

We have considered the case of a superfluid phase transition occurring
in the core of a pulsar and have studied observational effects of the
random vortex network formed via the Kibble-Zurek mechanism during the
transition.  Due to these vortices, extra angular momentum is generated
for the outer shell of the pulsar which can point in any arbitrary
direction. This leads to wobbling of the pulsar, along with possibility
of glitches/anti-glitches depending on the relative direction of this
extra angular momentum and the original angular momentum of the pulsar.
Importantly, the prediction of our model are almost universal, entirely
characterized by the symmetry breaking (U(1) in this case), and the
superfluid correlation length. Further, the observational effects on the
perturbations in the rotational dynamics of the pulsar should decay in
accordance with specific power law determined by power law decay of the
vortex network formed in the phase transition. These universal features
can help in identifying the nature of phase transition occurring inside
a pulsar core.

There are important aspects of our model which need further
investigations.  The decay of vortex network will lead to annihilation
of vortices/antivortices at the surface of $S^2$. This dynamics will be
complicated due to curved nature of the surface as vortices exiting in
different directions can slide on the surface and annihilate/join with
the vortices at different points. A proper study of this can only be done
by a detailed numerical simulation of the evolution of vortex network.
Even the calculation of net extra angular momentum pointing in certain
direction (with our assumption of applicability of Eq.(\ref{eq:sigma}) for 
different patches of the surface of $S^2$) needs to be verified with a 
detailed simulation. We plan to carry out such a simulation to address these
issues. We have ignored any discussion of the superconductivity 
arising from the Cooper pairing of protons in the neutron star core. All our 
arguments can be straightforwardly extended to this superconducting phase. 
Its implications will be very interesting as string defects will now be 
magnetic flux tubes, hence  extra vortices will imply changes in the 
magnetic field of neutron star. This will clearly have distinct observational 
signatures, and needs to be further explored.

\section{Acknowledgments}
We thank Shreyansh S. Dave and Arpan Das for useful discussions. 
Partha Bagchi acknowledges the financial support from Department of Atomic 
Energy (DAE) project RIN 4001.

\section{Data Availability}
No new data were generated or analysed in support of this research.

\bibliographystyle{mnras}
\bibliography{pulsarsf} 
\bsp
\label{lastpage}
\end{document}